\documentclass[12pt]{article}
\usepackage{latexsym}
\usepackage{amsmath,amssymb,amsfonts,amsthm}
\usepackage[english]{babel}
\newcommand{\be}{\begin{equation}}
\newcommand{\ee}{\end{equation}}
\newcommand{\bea}{\begin{eqnarray}}
\newcommand{\eea}{\end{eqnarray}}
\newcommand{\nn}{\nonumber \\}
\newcommand{\p}[1]{(\ref{#1})}
\newcommand{\lb}{\label}

\newcommand\s{\scriptscriptstyle}

\newcommand{\B}{{\s B}}

\newcommand\I{{\s I}}

\begin{document}

%%%%%%%%%%%%%%%%%%%%%%%%%%%%%%%%%%%%%%%%%%%%%%%%%%%%%%%%%%%%%%%%%%%%%%%%%
\begin{titlepage}

\begin{flushright}
ITP--UH--07/13
\end{flushright}
\vspace{0.5cm}

\begin{center}
\baselineskip=16pt {\Large\bf Auxiliary superfields}
\vspace{0.3cm}

{\Large\bf in ${\cal N}$=1  supersymmetric self-dual electrodynamics}
\vskip 1.5cm

{\large\bf
Evgeny Ivanov$\,^{1)}$, $\;$Olaf Lechtenfeld$\,^{2)}$, $\;$Boris Zupnik$\,^{1)}$ }
\vspace{1cm}

$^{1)}${\it Bogoliubov Laboratory of Theoretical Physics, JINR, \\
141980 Dubna, Moscow Region, Russia\\
}
\vspace{0.4cm}

$^{2)}${\it Institut f\"ur Theoretische Physik
and Riemann Center for Geometry and Physics, \\
Leibniz Universit\"at Hannover, \\
Appelstra{\ss}e 2, 30167 Hannover, Germany\\
}

\vspace{0.5cm}

{\tt eivanov,zupnik@theor.jinr.ru}, $\;$ {\tt lechtenf@itp.uni-hannover.de}

\end{center}
\vspace{1.3cm}

\par
\begin{center}
{\bf ABSTRACT}
\end{center}
\begin{quote}
We construct the general formulation of ${\cal N}{=}\,1$ supersymmetric
self-dual abelian gauge theory involving auxiliary chiral spinor superfields.
Self-duality in this context is just $U(N)$ invariance of the nonlinear
interaction of the auxiliary superfields. Focusing on the $U(1)$ case,
 we present the most general form of the $U(1)$ invariant auxiliary interaction,
 consider a few instructive examples
and show how to generate self-dual ${\cal N}{=}1$ models
with higher derivatives in this approach.

\vspace{1.5cm}


\end{quote}
\end{titlepage}

\setcounter{footnote}{0}

\setcounter{page}{1}
%%%%%%%%%%%%%%%%%%%%%%%%%%%%%%%%%%%%%%%%%%%%%%%%%%%%%%%%%%%%%%%%%%%%%%%%%%%%%%%%%

%\numberwithin{equation}{section}
%%%%%%%%%%%%%%%%%%%%%%%%%%%%%%%%%%%%%%%%%%%%%%%%%%%%%%%%%%%%%
\section{Introduction}

The study of duality-invariant (or self-dual) models of nonlinear electrodynamics
\cite{GZ}-\cite{AFZ} remains an active subject closely related to various issues
of current interest. Recently, the attention increased further, due to the
hypothetical crucial role of duality symmetries
in the possible ultraviolet finiteness of ${\cal N}{=}\,8$ supergravity and some
of its descendants (see, e.g., \cite{BHN}-\cite{BN}).

Some time ago, two of us \cite{IZ,IZ1} developed a new general formulation of
$U(1)$ duality-invariant models of nonlinear electrodynamics. This formulation
involves, besides the Maxwell gauge-field strength, some auxiliary bispinor
fields. The interaction in the full Lagrangian is constructed solely from
these auxiliary fields.  The standard nonlinear Lagrangian of any particular
duality-invariant system is recovered by eliminating these fields via their
equations of motion. The main advantage of the auxiliary bispinor formulation
is a linearization of the renowned self-duality condition~\cite{GZ}-\cite{GR}
in this setting. Self-duality becomes simply the requirement of off-shell
$U(1)$~
invariance of the interaction Lagrangian, and the $U(1)$~duality group gets
realized by linear transformations of the auxiliary fields.
The auxiliary bispinor approach admits an extension to $U(N)$ duality with
$N$~copies of the Maxwell field strength~\cite{IZ2} as well as to the
inclusion of additional scalar coset fields~\cite{prep}.

It was demonstrated in~\cite{IZ3} that the so called ``deformed twisted
self-duality condition'' recently proposed and exploited in~\cite{BN,CKR, CKO}
is in fact equivalent to the basic algebraic equation for the bispinor
auxiliary fields in the formulation of~\cite{IZ, IZ1}.

As suggested in \cite{IZ,IZ1,IZ3}, an obvious next step was to supersymmetrize
the auxiliary bispinor formulation, i.e.~to extend it to self-dual nonlinear
${\cal N}{=}1$ and ${\cal N}{=}2$ supergauge theories, e.g. starting from
the by now standard approach of~\cite{KT,KT2}. This step was recently accomplished
by Kuzenko~\cite{Ku},  who showed, in particular, that the auxiliary bispinor
field gets enhanced to a chiral fermionic ${\cal N}{=}1$ superfield.
In this language, the $U(1)$ duality amounts to manifest $U(1)$ invariance of
the auxiliary superfield interaction.

The purpose of the present paper is to introduce a framework which is more
general than the one given in~\cite{Ku}, concentrating on the rigid
${\cal N}{=}1$ case. Our invariant superfield density $E$ is a function of
three $U(1)$ invariant scalar superfield variables composed of
auxiliary chiral superfields $U_\alpha$. By analogy with \cite{IZ1,IZ3}
we also construct alternative formulations of ${\cal N}{=}1$ self-dual theories
which in addition make use of auxiliary scalar superfields. Another novelty of
our paper is a generalization of the ${\cal N}{=}1$ duality formulation with
auxiliary superfields from the abelian situation to the case of $U(N)$ duality.

The structure of the paper is as follows.
Section~2 recapitulates the basic features of self-dual ${\cal N}{=}1$ nonlinear
electrodynamics in the standard formulation.
In section~3 we outline the general formulation of ${\cal N}{=}1$ supersymmetric
$U(1)$ self-dual gauge theories, employing the auxiliary chiral (but otherwise
unconstrained) spinor superfield~$U_\alpha$ in parallel with the ordinary chiral
superfield strength~$W_\alpha\,$. We analyze the  equation of motion for
$U_\alpha(W,\bar{W})$ using a three-parametric $U(1)$ invariant superfield
density $E$ which is most general in the case without higher derivatives.
This equation is just ${\cal N}{=}1$ counterpart of the equation for the
auxiliary bispinor fields of the bosonic self-dual case, which was recently
rediscovered as the deformed twisted self-duality condition.
By eliminating  the auxiliary
superfield by a recursive procedure in terms of the ordinary covariant superfield
strengths $W_\alpha, \bar W_{\dot\alpha}$ we recover the standard representation
 of the general supersymmetric $U(1)$ self-dual theory.
In some particular parametrization, the interaction $E$ depends only on a single
 real variable, in a more direct analogy with the bosonic $U(1)$ self-dual
 theories~\cite{IZ1} (this case was treated in~\cite{Ku}).
In section 4 we present an alternative self-dual ``$M$~representation'' which
makes use of an additional scalar superfield~$M$.
It is a supersymmetric extension of the so called ``$\mu$ representation''
 of the bosonic case \cite{IZ1,IZ3}. Like its bosonic prototype,
the $M$ representation is capable of essentially simplifying the calculations.
Examples of $U(1)$ self-dual theories are studied in section~5.
We translate to our formulations the renowned ${\cal N}{=}1$ Born-Infeld theory
as well as construct a new self-dual model specified by a simple quartic
interaction of the auxiliary superfields. It is the ${\cal N}{=}1$ extension of
the quartic auxiliary bosonic interaction considered firstly in~\cite{IZ,IZ1}
and recently discussed in~\cite{BN,CKR}. We present, as a perturbative expansion,
the relevant superfield actions in terms of the ordinary superfield strengths.
We also show how to adjust our approach for constructing supersymmetric
self-dual models containing higher derivatives and
give a few examples of such models. The bosonic limit of the formulation with
auxiliary spinor superfields is discussed in section~6.
We give the bosonic component actions for a few examples, including those with
higher derivatives.
A brief account of the generalization of the new self-duality setting to the
$U(N)$ case is the subject of section~7.
We present several examples of $U(N)$ self-dual models which correspond to some
particular choices of the invariant auxiliary interaction.

\section{Nonlinear ${\cal N}{=}1$  electrodynamics}
Here we fix our notations and sketch the superfield formalism of the
${\cal N}{=}1$
self-dual theories \cite{CF,BG,KT,KT2}.

Our conventions are the same as, e.g., in the book \cite{WB} and in Refs.
\cite{KT,KT2}.
We parametrize the ${\cal N}{=}1, D{=}4$ superspace  by the coordinates
\be
z=(x^m, \theta^\alpha,
\bar\theta^{\dot\alpha}),\quad \theta^2=\theta^\alpha
\theta_\alpha,\quad\bar\theta^2=
\bar\theta_{\dot\alpha}\bar\theta^{\dot\alpha}
\ee
and define the covariant spinor derivatives as
\bea
&&D_\alpha=\partial_\alpha+i\bar\theta^{\dot\alpha}(\sigma^m)_{\alpha\dot\alpha}
\partial_m,\qquad
D^2=D^\alpha D_\alpha,\nn
&&\bar{D}_{\dot\alpha}=-\bar\partial_{\dot\alpha}- i\theta^\alpha
(\sigma^m)_{\alpha\dot\alpha}
\partial_m,\quad\bar{D}^2=\bar{D}_{\dot\alpha}\bar{D}^{\dot\alpha}\,.
\eea
Here $\alpha$ and $\dot\alpha$ are the doublet $SL(2,C)$ indices and
$m=0,1,2,3$ (we use the flat Minkowski metric $\eta_{mn} = {\rm diag} (1, -1,-1,-1)\,$).

The Grassmann integrals are normalized as
\bea
&&\int d^2\theta \theta^2=-\frac14D^2 \theta^2=1~,\quad \int
d^2\bar\theta \bar\theta^2
=-\frac14\bar{D}^2 \bar\theta^2=1~,\\
&&\int d^4xd^2\theta \equiv\int d^6\zeta ~,\quad\int d^4xd^2\theta
d^2\bar\theta\equiv\int d^8z.
\eea

The (anti)chiral Abelian superfield strengths are defined by
\bea
W_\alpha=-{1\over4}\bar{D}^2A_\alpha=-\frac14\bar{D}^2D_\alpha
V~,\quad \bar W_{\dot\alpha}=-\frac14D^2\bar{A}_{\dot\alpha}=
-\frac14D^2\bar{D}_{\dot\alpha} V\,,
\eea
where $V$ is the gauge
prepotential and $A_\alpha = D_\alpha V\,, \; \bar{A}_{\dot\alpha} = \bar{D}_{\dot\alpha} V$ are
spinor gauge
connections. The superfield strengths satisfy, besides the chirality conditions
\be
\bar{D}_{\dot\alpha}W_\beta = 0\,, \quad D_\alpha  \bar W_{\dot\alpha} = 0\,, \lb{chirW}
\ee
also the Bianchi identity:
\be
B(W,\bar W)\equiv D^\alpha
W_\alpha-\bar{D}_{\dot\alpha} \bar{W}^{\dot\alpha}
=0~. \lb{Bianchi}
\ee

For what follows, it will be useful to tabulate the $R$ invariance properties
of various ${\cal N}{=}1$
quantities:
\bea
&&R(\theta^\alpha)=1,\quad R(\bar\theta^{\dot\alpha})=-1,\quad
R(D_\alpha)=-1,\quad R(\bar{D}_{\dot\alpha})=1,\nn
&&R(W_\alpha)=1,\quad R(\bar{W}_{\dot\alpha})=-1.
\eea

The ``engineering'' dimensions of the basic objects of the ${\cal N}{=}1$ gauge
theory are (in the
mass units):
\be
[V] = 0\,, \; [W_\alpha] = 3/2\,, \;  [D_\beta W_\alpha] = 2\,,
\ee
and the free superfield action is written as
\be
S_2(W,\bar{W})=\frac1{4}\int d^6\zeta W^2 +\frac1{4}\int d^6\bar\zeta \bar
W^2\,,\lb{free}
\ee
where $W^2\equiv W^\alpha W_\alpha$ and $\bar W^2\equiv\bar W_{\dot\alpha}
\bar W^{\dot\alpha}$.

In the nonlinear theory with one dimensionful constant $f$ ($[f] = -2$),
it is convenient to ascribe nonstandard
dimensions to the basic objects,
\be
[V]= -2~,\qquad
[W_\alpha]= -1/2~,\qquad [D_\alpha W_\beta]=  0\,,
 \ee
and to construct the nonlinear action as
$$
S = f^{-2}[S_2(W,\bar{W})+S_{int}(W,\bar{W})]\,,
$$
where
\be
S_{int}(W,\bar{W}) = \frac14 \int d^8 z\, {\cal L}_{int}(W, \bar W)\,.\lb{intW}
\ee
For convenience, we will put $f=1$ altogether.

We consider the following form of the arbitrary nonlinear interaction
in the $W$ representation :
\be
{\cal L}_{int}= W^2\bar{W}^2\Lambda(w,\bar{w}, y, \bar{y})\,,\lb{Lambda}
\ee
where the superfield density $\Lambda$
depends on the dimensionless R-invariant variables
\bea
&&w=\frac18\bar D^2\bar{W}^2~,\quad \bar{w}=\frac18D^2W^2~,\qquad \quad
y\equiv D^\alpha
W_\alpha\,. \lb{wbarwy}
\eea
A wide subclass of nonlinear models (including ${\cal N}{=}1$ super Born-Infeld
theory)
is associated with the $y$-independent densities $\Lambda(w,\bar{w})\,$.
In \cite{KT} just
this set of models
was mainly addressed.

Let us define
\bea
&&M_\alpha\equiv-2i\frac{\delta S}{\delta W^\alpha},\qquad
 \bar{M}_{\dot\alpha}\equiv 2i\frac{\delta S}{\delta
\bar{W}^{\dot\alpha}}.
\eea
Then the nonlinear equations of motion can be written in the form
\footnote{While computing the
variations of the action, one should treat $W_\alpha, \bar W_{\dot\alpha}$
as unconstrained chiral
superfields which are not subjected to the Bianchi identity \p{Bianchi}.
Correspondingly, the variables
$y$ and $\bar y$ are considered as
independent. The Bianchi identity $y = \bar y$ is imposed {\it \`a posteriori}.
The equations of
motion \p{nonleq} expressed
through the so defined $M_\alpha$ and $\bar M_{\dot\alpha}$ coincide with those
derived by varying
$S$ with respect to the prepotential $V\,$.}
\be
N(W,\bar W)\equiv D^\alpha M_\alpha-\bar{D}_{\dot\alpha}
\bar{M}^{\dot\alpha}=0\,.\lb{nonleq}
\ee
It is straightforward to find the explicit expression for the chiral
superfield $M_\alpha$
\bea
M_\alpha= -iW_\alpha\left[1 -\frac{1}{4}\bar{D}^2\left\{\bar{W}^2[
\Lambda+\frac{1}{8}D^2(W^2 \Lambda_{\bar{w}})]\right\} \right]
-\frac{i}{8}\bar{D}^2\left[\bar{W}^2
D_\alpha(W^2 \Lambda_y)\right].\lb{Mrel0}
\eea

The $O(2)$ duality transformations defined as
\be
\delta W_\alpha=\omega
M_\alpha(W,\bar W)~,\quad \delta  M_\alpha=-\omega W_\alpha\,, \lb{N1duality}
\ee
mix the equation of motion \p{nonleq} with the
Bianchi identity \p{Bianchi}, leaving their set covariant.
The  $O(2)$ self-duality constraint ensuring the compatibility of \p{N1duality}
with the expression
\p{Mrel} for $M_\alpha$
and generalizing the bosonic Gaillard - Zumino condition has, in the
present case,  the integral
form \cite{KT}:
\be
\mbox{Im}\,K(W,\bar W)=0\,,\lb{n1self}
\ee
\be
K(W,\bar{W}) := -\int d^6\zeta\,(W^2 +M^2)\,.
\lb{n1chir}
\ee
The funcitonal $K(W,\bar W)$ in itself is invariant under \p{N1duality}.
In view of the nilpotency property $W_\alpha W_\beta W_\gamma = 0$, for
calculating $K(W, \bar{W})$
it is sufficient to know $M_\alpha$ in \p{Mrel} only up to terms linear in
$W_\alpha$:
\bea
M_\alpha|= -iW_\alpha\left[1 -\frac{1}{4}\bar{D}^2\left(\bar{W}^2\,\Gamma\right)
 \right] +\frac{i}{4}W^\beta\bar{D}^2\left(\bar{W}^2D_\alpha W_\beta \Lambda_y
 \right),\lb{Mrel}
\eea
where
\bea
\Gamma=\Lambda+\bar{w}\Lambda_{\bar{w}}=
\frac{\partial (\bar{w}\Lambda)}{\partial \bar{w}}\,.
\eea
Then, using once more the nilpotency property, this time for
$\bar W_{\dot\alpha}$, we can
write the functional $K(W, \bar W)$ as
an integral over the full ${\cal N}=1$ superspace
\bea
K(W,\bar W)=2\int d^8z\,W^2\bar{W}^2\left[\Gamma-w\Gamma^2 +2w \bar w\,
(\Lambda_y)^2\right].\lb{Kwy}
\eea
All relations are simplified if $\Lambda_y=0$. In this case
\bea
K(W,\bar W)=2\int d^8z\,W^2\bar{W}^2\,(\Gamma-w\Gamma^2)\,.
\eea

The notorious example of the ${\cal N}{=}1$ self-dual system with $\Lambda_y=0$
is the ${\cal N}{=}1$
generalization
of the BI theory \cite{CF,BG}. The function $\Lambda$ in this case is known
in a closed form:
%The first few terms in  its $w, \bar w$ expansion are as follows
\bea
&&\Lambda_{\s BI}
=\left[1+{1\over2}(w+\bar{w})+\sqrt{1+(w+\bar{w})
+{1\over4}(w-\bar{w})^2}\right]^{-1}\lb{N1BI}\\&&
={1\over2}-{1\over4}(w+\bar{w})+{1\over8}(w+\bar{w})^2+{1\over8}w\bar{w}
+\ldots\nonumber
\eea
By some rather tedious work one can check the validity of the self-duality
condition \p{n1self}
for $\Lambda_{\s BI}\,$.
%It is not valid for any other version of the
%${\cal N}{=}1$ BI theory described in \cite{CF},
%only this version proves to be self-dual.

The ${\cal N}{=}1$ BI action can be also written in the concise
form as
\be
S_{\s BI}=\frac{1}{4}\int d^6\zeta X+\mbox{c.c.}\,, \lb{BIX}
\ee
where the auxiliary chiral superfield $X$ satisfies the quadratic constraint
\cite{BG}
\be X+\frac1{16}X\bar{D}^2\bar X=W^2.\lb{BIquad}
\ee
The action $S_{\s BI}$ is also distinguished in that it is invariant under the
nonlinearly
realized second ${\cal N}{=}1$ supersymmetry which extends the manifest
${\cal N}{=}1$ supersymmetry
to ${\cal N}{=}2\,$ \cite{BG,RT}.

Another possible parametrization of the interaction density in \p{Lambda} is
through the variables
\bea
w^\prime=w+\frac18\bar{y}^2\,, \;\bar w^\prime= \bar w+\frac18 {y}^2\,, \; y,
\bar y\,, \lb{newPar}
\eea
so that
\be
\Lambda(w,\bar{w},y,\bar{y}) = \Lambda^\prime(w^\prime,\bar{w}^\prime,y,
\bar{y})\,.
\ee
The invariant functional \p{Kwy} and the self-duality condition can be rewritten
in the new parametrization with the help of
 the relations
\be
\frac{\partial\Lambda}{\partial w}=\frac{\partial\Lambda'}{\partial w'},
\quad
\frac{\partial\Lambda}{\partial y}=\frac{\partial\Lambda'}{\partial y}+
\frac14y\frac{\partial\Lambda'}{\partial \bar{w}'}\,.
\ee

\setcounter{equation}0
\section{${\cal N}=1$ self-duality with auxiliary chiral spinors}
\subsection{New representation for nonlinear ${\cal N}=1$ electrodynamics:
general setting}
We introduce the auxiliary chiral spinor superfield $U_\alpha$
and construct the following quadratic action:
\bea
&&S_2(W,U)=\int d^6\zeta\left(U W-{1\over2}U^2-{1\over4}W^2\right)+
\mbox{c.c.}\,,
\lb{freeWV}
\eea
where $U^2=U^\alpha U_\alpha\,, \bar{U}^2=\bar{U}_{\dot\alpha}
\bar{U}^{\dot\alpha}$.
Integrating out the auxiliary superfield,
\be
U_\alpha=W_\alpha\,,\lb{auxfree}
\ee
we reproduce the standard ${\cal N}{=}1$ quadratic action \p{free}.
The action \p{freeWV} is ${\cal N}{=}1$ analog of the free Maxwell action
rewritten through
the auxiliary bispinor fields \cite{IZ,IZ1}. This modified Maxwell action is
just the bosonic core
of \p{freeWV} (modulo auxiliary fields vanishing
on shell, see section 6).

The full set of equations of motion associated with  \p{freeWV} (including
\p{auxfree}), together
with the Bianchi identity \p{Bianchi}, is covariant
under the following $U(1)$ duality transformations
\bea
&& \delta U_\alpha=-i\omega U_\alpha~,\quad \delta W_\alpha =
i\omega( W_\alpha- 2 U_\alpha) \equiv
\omega M_\alpha(U, W),
\quad \mbox{and c.c.} \,, \lb{dual1} \\
&&\delta (W_\alpha-U_\alpha)=i\omega(W_\alpha-U_\alpha), \quad \mbox{and c.c.}
\,.\lb{dual2}
\eea
Though in the free case, after substituting \p{auxfree},  $M_\alpha (U,W)$
becomes just $iW_\alpha\,$, we will assume
that the same $U(1)$ transformations \p{dual1} act as the duality ones in the
general interaction case too, when \p{auxfree}
is replaced by a nonlinear equation and
$M_\alpha (U,W) = i( W_\alpha- 2 U_\alpha)$ becomes a nontrivial functional
of $W_\alpha, \bar W_{\dot\alpha}\,$.

The most general interaction of the auxiliary superfield, before imposing
any self-duality constraint, can
be chosen as
\be
S_E(U)=\frac{1}{4}\int d^8z\, U^2\bar U^2\,E(u,\bar{u}, g,\bar{g})\,,\lb{intaux}
\ee
where $E$ is an arbitrary real function  of the dimensionless Lorentz invariant
superfield variables
\be
u=\frac18\bar{D}^2\bar{U}^2, \quad \bar{u}=\frac18D^2U^2, \quad
g=D^\alpha U_\alpha,
\quad
\bar{g}=\bar{D}_{\dot\alpha}\bar{U}^{\dot\alpha}.
\ee
So the total action is
\be
S_{tot} = S_2(W,U) + S_E(U)\,.\lb{totWV}
\ee

The counterpart of the free auxiliary equation \p{auxfree} is obtained by
varying \p{totWV} with respect
to $U_\alpha$
\bea
W_\alpha-U_\alpha=-\frac{\delta S_E}{\delta U^\alpha}=
\frac18U_\alpha\bar{D}^2\{\bar{U}^2[ E+\frac18D^2(U^2 E_{\bar{u}})]\}
-\frac{1}{16}\bar{D}^2[\bar{U}^2D_\alpha(U^2 E_g)]\,.
\lb{WV}
\eea
Varying the full action with respect to the prepotential $V$, we obtain
the dynamical equation
\bea
D^\alpha( W_\alpha- 2U_\alpha)+ \bar{D}_{\dot\alpha}(\bar{W}^{\dot\alpha}
-2\bar{U}^{\dot\alpha})= 0\,.
\lb{nonlWU}
\eea
Comparing it with \p{nonleq}, we identify, as in the free case,
\be
W_\alpha- 2U_\alpha = -iM_{\alpha}(W,U)\,, \quad \bar W_{\dot\alpha}-
2\bar U_{\dot\alpha} =
i\bar M_{\dot\alpha}(W,U)\,.\lb{WUequ}
\ee
Using the algebraic equation \p{WV} and substituting its solution
$U_\alpha = U_\alpha (W,\bar W)$ into \p{nonlWU},
we reproduce the initial form of the nonlinear equation of motion \p{nonleq}.

Note that in practice, while solving \p{WV} to restore the $W, \bar W$
representation of the total superfield action
$S_{tot}\,$, it is enough to consider
an effective form of \p{WV}, in which only terms linear in $U_\alpha$ are kept
\bea
W_\alpha| =U_\alpha + \frac18U_\alpha\bar{D}^2\,[\bar{U}^2( E+
\bar uE_{\bar{u}})]
+\frac{1}{8}U^\beta\bar{D}^2(\bar{U}^2D_\alpha U_\beta E_g).
\lb{WV1}
\eea
This is due to the property that in \p{freeWV} all interaction terms appear
at least with the factor $W^2$
(or with $\bar W^2$ in the complex conjugated part), while in \p{intaux}
with the factor $W^2\bar W^2\,$.
It will be also useful to rewrite \p{WV1} as
\bea
W_\alpha| = U_\alpha + C_\alpha^\beta\,U_\beta\,, \quad
C_\alpha^\beta = \frac18\bar{D}^2 (\bar U^2 R_\alpha^\beta)\,,\quad
R_\alpha^\beta := \,[\delta^\beta_\alpha (E + \bar u E_{\bar u}) -
D_\alpha U^\beta \,E_g]\,. \lb{CRmatrix}
\eea
Note that, up to the nilpotent terms $\sim \bar U_{\dot\alpha}$,
\be
C_\alpha^\beta = u R^\beta_\alpha + O(\bar U)\,,
\lb{Rdef}
\ee
so the matrix $C_\alpha^\beta$, when multiplied by $\bar U^2$, is reduced to
its ``effective'' form  $uR^\beta_\alpha\,$.

\subsection{Relation to the original formulation}
The general representation for the perturbative solution of the Eq.\p{WV1},
under the assumption that all nilpotent terms can be ignored,
has the  form
\bea
&&U_\alpha(W,\bar W)\approx [\delta^\beta_\alpha +
{\cal B}_\alpha^\beta(W,\bar W)] W_\beta\,.\lb{Usol}
\eea
The chiral dimensionless
matrix function ${\cal B}_\alpha^\beta(W,\bar{W})$ satisfies the relation
\bea
&&(\delta^\beta_\alpha +{\cal B}_\alpha^\beta)(\delta_\beta^\gamma
+C_\beta^\gamma)=\delta^\gamma_\alpha.
\eea
It can be parametrized by two superfields $G(W,\bar W)$ and $ P(W, \bar W)$
\bea
&&{\cal B}_\alpha^\beta\approx
 u[G\,\delta_\alpha^\beta+ P\,D_\alpha W^\beta]\,.
\eea
This representation is analogous to the one used in the bosonic case \cite{IZ1}
\bea
V_{\alpha\beta}=G(F^2)F_{\alpha\beta},\quad G(F^2)=\frac12
-\frac{\partial L}{\partial (F^2)}=
\left[1+\frac{\partial E}{\partial (V^2)}\right]^{-1},
\eea
where $L(F)$ is the nonlinear Maxwell Lagrangian, and $E(V^2,\bar{V}^2)$
is the auxiliary interaction.

Expressing $U_\alpha$ through $M_{\alpha}$ from \p{WUequ}, we can  directly
compare the solution
\p{Usol} with the formula \p{Mrel} in the
original $W$ representation
\bea
U_\alpha=\frac12W_\alpha+\frac{i}2M_\alpha =W_\alpha\left[1-\frac{1}{8}
\bar{D}^2[\bar{W}^2(\Lambda+\bar{w}\Lambda_{\bar{w}})]\right]
+\frac{1}8W_\beta\bar{D}^2(\bar{W}^2D_\alpha W^\beta \Lambda_y)\,.\lb{UWW}
\eea
Thus derivatives of the interaction density $\Lambda$ in the $W$ representation
can be found directly from the solution \p{Usol} in analogy with the bosonic
formalism.
Examples of such calculations are considered in section 4.

Now let us derive the relation between the interaction functions in the
$W$ and $W,U$ representations, i.e. between $\Lambda (w, \bar w, y, \bar y)$
and $E(u, \bar u, g, \bar g)$.

As the first step, we use Eqs. \p{Rdef} to find the effective relation between
$W^2$ and $U^2$,
\bea
W^2| &=& U^2\Big[1 +C_\sigma^\sigma
-\frac12C_\rho^\sigma C_\sigma^\rho + \frac12C_\sigma^\sigma C_\rho^\rho\Big] \nn
&=& U^2\Big\{1 + \frac18\bar D^2\,[\bar U^2 (R_\sigma^\sigma
- \frac12 C_\rho^\sigma\,R^\rho_\sigma +  \frac12 C_\sigma^\sigma\,R^\rho_\rho)]
\Big\}.  \lb{W2}
\eea
This expression, modulo the nilpotent terms $\sim \bar U_{\dot\alpha}\,$,
can be rewritten as
\bea
W^2| = U^2\, H\,, \quad H = [1+u( E+\bar{u} E_{\bar{u}})]^2
+[1+u( E+\bar{u} E_{\bar{u}})]u g E_g
-2\bar{u}u^2E_g^2\,. \lb{Pden}
\eea
It allows to find the exact relation between the 4-th order nilpotent terms
\bea
W^2\bar{W}^2 = U^2\bar{U}^2\, H(u,\bar{u},g,\bar{g})\bar H(u,\bar{u},g,\bar{g})\,.
\lb{w2w2}
\eea

As the next step, we represent $S_2(W,U)\,$ \p{freeWV} as
$$
S_2(W,U) = \int d^6\zeta\Big[\frac14 W^2 - \frac12(U -W)^2\Big] +\mbox{c.c.}\,.
$$
Using Eq.\p{WV} and the relations
\bea
D_\alpha U^\beta\, D^\alpha U_\beta\,U^2 =  4 \bar u\,U^2\,\,, \quad
D_\alpha U^\beta\,D_\beta U^\alpha\,U^2
= (4 \bar u+g^2)\,U^2\,, \lb{relUU}
\eea
we find
\bea
(W-U)^2=\frac12U^2(C_\sigma^\sigma C_\rho^\rho -C_\rho^\sigma C_\sigma^\rho),
\eea
and
\bea
\int d^6\zeta\, (U-W)^2
= -\frac1{2}\int d^8zU^2\bar{U}^2T(u,\bar{u},g,\bar{g}),
\eea
where
\bea
T(u,\bar{u},g,\bar{g})=u( E+ \bar{u} E_{\bar{u}})^2
+ ug E_g( E+ \bar{u} E_{\bar{u}})
- 2u {\bar{u}} E_g^2.\lb{Tden}
\eea

Finally, we equate the actions $S_{tot}$ in the $W$ and $W,U$ representations
and obtain
\bea
\Lambda=\frac{1}{H\bar H}\Big(E +T+\bar{T}\Big)\,, \lb{WWUlagr}
\eea
where $H$ and $T$ are the nonlinear
combinations
of $E, E_u, E_{\bar{u}}, E_g$
and $E_{\bar{g}}$ defined in Eqs. \p{Pden} and \p{Tden}.

To find the explicit form of $\Lambda$ we also need to express the superfield
arguments $u, \bar u, d, \bar d$
in the r.h.s. of \p{WWUlagr}
in terms of the original variables $w, \bar w, y, \bar y\,$. Because
$\Lambda$ appears in the action with
the factor $W^2\bar W^2$, we can use the
effective version of the equations relating these two sets of variables
\bea
&&w| = u \,\bar H\,, \qquad \bar w| = \bar u \,H\,,\lb{wueff}\\
&&D_\beta W_\alpha|=D_\beta U_\rho(\delta^\rho_\alpha+uR_\alpha^\rho),\nn
&&
y|=g+ u\,D^\alpha U_\rho R_\alpha^\rho=g+2ug(E+\bar{u}E_{\bar{u}})
-4u\bar{u}E_g\,.\lb{DWeff}
\eea
When solving these effective equations, one can exploit the relations
analogous to \p{relUU}
\bea
D_\alpha W^\beta D^\alpha W_\beta\,W^2 =  4 \bar w\,W^2\,\,, \quad
D_\alpha W^\beta\,D_\beta W^\alpha\,W^2
= (4 \bar w +y^2)\,W^2\,. \lb{relWW}
\eea

\subsection{${\cal N}=1$ self-duality condition in the $(W,U)$ representation}
Now we assume that the $U(1)$ duality transformations \p{dual1}, \p{dual2}
retain their form in the case
with interaction (by analogy with
the bosonic case) and wish to learn how the supersymmetric $U(1)$ duality
constraint \p{n1self}, \p{n1chir} looks in the
formulation  with the auxiliary spinor superfields.

The $U(1)$ invariant \p{n1chir} in the formulation considered becomes
\bea
K=- \int  d^6\zeta(W^\alpha W_\alpha+M^\alpha M_\alpha)=
-4\int  d^6\zeta[U^\alpha (W_\alpha-U_\alpha)]
= -4\int  d^6\zeta U^\alpha \frac{\delta S_E}{\delta U^\alpha}\,.
\eea
Thus the supersymmetric $U(1)$  self-duality constraint  \p{n1self} in
this formulation is none other
than the condition of $U(1)$
invariance of the auxiliary interaction \p{intaux}:
\bea
K-\bar{K}=0~\rightarrow~\delta_\omega S_E=0\,.
\eea
Now it is easy to check that, for such $U(1)$ invariant self-interactions
$S_E$, the equation \p{WUequ},
together with the dynamical equation
\p{nonlWU} and the Bianchi identities \p{Bianchi}, are covariant under
the transformations \p{dual1}, \p{dual2}.

We conclude that  the whole family of self-dual models of the nonlinear
${\cal N}{=}1$ supersymmetric
electrodynamics
is parametrized by $U(1)$ invariant superfunction
$E_{inv}(u, \bar u, g, \bar g)\,$.

{}From the variables $u, \bar u, g, \bar g$ one can construct four
$U(1)$ invariants,
\bea
A := u\bar u\,, \quad C :=g\bar{g}\,,\quad B := ug^2\,, \quad
\bar{B} :=\bar{u}\bar{g^2}\,,
\eea
which are connected by the relation
\be
B\bar B = A C^2\,. \lb{relADB}
\ee
We are interested in such interactions which are analytic at the point
$A=C=B=0$ and so admit power
series expansion in these $U(1)$
invariant variables. With making use of the relation \p{relADB}, it is easy
to show that the most
general $U(1)$ invariant self-interaction of this
type is given by the following ansatz
\be
E_{inv} ={\cal F}(B, A, C)+\bar{\cal F}(\bar{B}, A, C)\,, \lb{Einv}
\ee
where ${\cal F}(B,A,C)$ is an arbitrary analytic function.
 This is in contrast with the
pure bosonic $U(1)$ self-dual systems
which are parametrized by a real function ${\cal E}$ which depends on only
one real variable
$a=V^2\bar{V}^2\,, \; V^2 := V^{\alpha\beta}V_{\alpha\beta}\,, \; \bar V^2 :=
\bar V^{\dot\alpha\dot\beta}V_{\dot\alpha\dot\beta}$ \cite{IZ}. The expansions
of
${\cal F}(B, A, C)$ and $\bar{\cal F}(\bar{B}, A, C)$ as formal series with
constant coefficients look like
\bea
&&{\cal F}=e_1+e_2A+f_1C+f_2C^2+h_1B+O(U^6)\,,\nn
&& \bar{\cal F}=e_1+e_2A+f_1C+f_2C^2+\bar{h}_1\bar{B}
+O(U^6)\,.
\eea

The derivatives of $E_{inv}$ can be rewritten through the independent ones
as follows
\footnote{For brevity, we omit the index `$inv$' on $E$.}
\bea
&&E_{\bar{u}}=u{\cal F}_A+u\bar{\cal F}_A+\bar{g}^2\bar{\cal F}_{\bar{B}},\quad
\bar{u}E_{\bar{u}}=AE_A+\bar{B}\bar{\cal F}_{\bar{B}},\nn
&&E_g=\bar{g}{\cal F}_C+\bar{g}\bar{\cal F}_C+2ug{\cal F}_B,\quad
gE_g=CE_C+2B{\cal F}_B.\lb{invder}
\eea
Using these formulas, all the general relations and quantities, including
the matrices $C_\alpha^\beta$
and $R_\alpha^\beta$
defined in \p{CRmatrix},  can be easily specialized to the $U(1)$ invariant case. In particular, the
relation \p{WWUlagr}
for the duality-invariant case is obtained by
replacing there $E \rightarrow E_{inv} = {\cal F} + \bar{\cal F}$ and
expressing the derivatives with
respect to $u, \bar u, g, \bar g$
in terms of ${\cal F}_A,  {\cal F}_C$, ${\cal F}_B$ (and their
complex-conjugates) according to
the formulas \p{invder}.

We also note that sometimes it is more convenient to use the equivalent set
of the superfield variables (cf. Eq. \p{newPar})
\be
u' = u  + \frac{1}{8}\,\bar g^2\,, \quad \bar u' = \bar u  +
\frac{1}{8}\,g^2\,, \quad g\,, \quad \bar g\,,
 \lb{prime}
\ee
and, respectively,
\bea
E(u, g)=E'(u',g),\quad \frac{\partial E}{\partial u}=
\frac{\partial E'}{\partial u'},\quad
\frac{\partial E}{\partial g}=\frac{\partial E'}{\partial g}+
\frac14g\frac{\partial E'}{\partial \bar{u}'}\,.
\eea
The new invariant variables in the self-dual theories have the form
\be
A' = u'\bar u'\,, \quad C = g\bar g\,, \quad B' = u' g^2\,, \quad
\bar B' = \bar u' \bar g^2\,. \lb{prime2}
\ee
They have an advantage of possessing a simpler component expansion.

\subsection{A particular subclass of ${\cal N}=1$ self-dual interactions}

Now we consider the particular choice of the $U(1)$ invariant interaction
$E(A)$ involving only one real
superfield variable
\be
A=u\bar{u}=\frac1{64}(D^2U^2)(\bar{D}^2\bar{U}^2).
\ee
Just this case was treated in \cite{Ku}. The auxiliary equation \p{WV} is
reduced to
\bea
&&W_\alpha-U_\alpha=\frac{1}{8}U_\alpha
\bar{D}^2\{\bar{U}^2[ E+ \frac18 D^2\left(U^2
u\,E_A\right)]\}\lb{1paramet1}
\eea
and the effective equations take the form
\bea
&&W_\alpha| =U_\alpha[1+u {\cal P}(A)],\quad {\cal P}(A)=\frac{d}{dA}(A E)=
 E+A E_A,\\
&&\bar{w}| =\bar{u}[1+u {\cal P}(A)]^2\,, \quad {w}| ={u}[1+
\bar u {\cal P}(A)]^2\,.\lb{WUpart}
\eea
These effective equations are analogous to the bosonic equations of
Ref. \cite{IZ1}
\bea
F_{\alpha\beta}=V_{\alpha\beta}(1+\bar{V}^2{\cal E}_a),\quad F^2=V^2
(1+\bar{V}^2{\cal E}_a)^2\,,
\eea
where ${\cal E} = {\cal E}(a)\,, \;a = V^2\bar V^2\,$. The similarity is
based on the formal correspondence
\be
u\,\leftrightarrow \,\bar{V}^2,\quad A\, \leftrightarrow  \,a,\quad
{\cal P}(A)\, \leftrightarrow \,
{\cal E}_a,
\ee
which can in fact be trusted by the component consideration (see Sect. 6).

The relation \p{WWUlagr} in the present case reads
\be
\Lambda = \frac{ E + (u + \bar u){\cal P}^2(A)}{[1+u {\cal P}(A)]^2
[1+ \bar u {\cal P}(A)]^2}\,.
\ee

It is instructive to give few first terms in the power series expansion
of $\Lambda$ in the $W$ representation,
starting from ${\cal P}(A)=e_1+e_2A+\ldots\,, \;
E(A) = e_1 + \frac12e_2A+\ldots\,$.

For $U_\alpha$ we obtain the recursive equation
\bea
&&U_\alpha=W_\alpha\,\frac1{1+u{\cal P}}=W_\alpha G(w,\bar{w})\nn
&&=W_\alpha[1-e_1u-e_2u^2\bar{u}+e^2_1u^2 - e_1^3u^3+O(U^{8})],
\eea
which implies
\bea
U_\alpha = W_\alpha\left[1 - e_1w + e_1^2(w^2 + 2w\bar w) - e_1^3(w^3
+ 3 w\bar w^2 + 8 w^2\bar w) -
e_2 w^2\bar w + O(W^8)\right]\lb{UWexpr}
\eea
and
\bea
u = w - 2e_1 w\bar w + e_1^2\,w\bar w (4 w  + 3 \bar w) + O(W^8)\,.\lb{uwspec}
\eea
For $\Lambda$ we have the following $(u, \bar u)$ expansion
\bea
\Lambda &=& e_1 - e_1^2 (u + \bar u) + [e_1^3(u^2 + \bar u^2) +
\frac12 e_2 u\bar u]
+ (e_1^4 -e_1e_2)(u\bar u^2  + u^2\bar u)\nn
&& -\, e_1^4(u^3 + \bar u^3) + O(U^8)\,,
\eea
which, after substituting \p{uwspec}, yields
\bea
\Lambda &=& e_1 - e_1^2(w + \bar w) + (4e_1^3 + \frac12 e_2)w\bar w +
e_1^3 (w^2 + \bar w^2)\nn
&&- \,2(e_1e_2 + 5 e_1^4)(w \bar w^2 + w^2\bar w)
 - e_1^4(w^3 + \bar w^3) + O(W^8)\,.
\eea
This perturbative solution for $\Lambda$, together with the expression
\p{UWexpr}, nicely agree with
the effective form of
Eq. \p{UWW} for the considered case,
\bea
&&U_\alpha|=W_\alpha\left[1-w(\Lambda+\bar{w}\Lambda_{\bar{w}})\right].\lb{Usol2}
\eea

\setcounter{equation}{0}
\section{Alternative auxiliary superfield representation}

Here we construct ${\cal N}{=}1$ analog of the so called $\mu$
representation of the bosonic case. We term it ``$M$ representation''. It
seemingly exists only for the subclass of self-dual theories considered in
\cite{Ku}.

Besides the chiral spinor superfields $W_\alpha$ and $U_\alpha$ we introduce a
complex general scalar
${\cal N}=1$ superfield $M$ and
construct the following ``master'' action
\bea
S_{mast} = S_2(W,U) + S_{int}(W, U, M)\,, \lb{mast}
\eea
where $S_2(W,U)$ is the same as in \p{freeWV} and
\bea
S_{int}(W, U, M) = \frac14 \int d^8z\left[(U^2\bar M + \bar U^2 M) +
M\bar M\,J\,(m, \bar m) \right],
\lb{intmast}
\eea
with
\be
m = \frac18 \bar D^2 \bar M\,, \quad \bar m  = \frac18 D^2 M\,.
\ee
The interaction function $J(m, \bar m)$ is real. For the special choice of the
interaction function,
$J_{inv} = J_{inv}(B)\,, \;B := m\bar m\,$,
the action \p{intmast}  is invariant under the duality $U(1)$ transformations
realized on the newly
introduced superfield $M$ as
\be
\delta M = 2i\omega M\,, \quad \delta \bar M = -2i \omega \bar M\,,\quad
\delta\bar{m}=2i\omega\bar{m}\,,\quad \delta m=2i\omega m\,.\lb{Ntrans}
\ee

\subsection{From master action to the $(W,U)$ formulation}
Let us firstly show that, eliminating the auxiliary superfields $M\,, \bar M$,
we will recover the particular case of the
action \p{intaux} with $E = E(u, \bar u)\,$. We assume that $J$ starts with
a constant, so the function $J^{-1}$ is well defined
at the origin.

The corresponding equations of motion are
\bea
M = -J^{-1}\left[U^2 + \bar D^2(M\bar{M} J_m) \right], \quad
\bar M = -J^{-1}\left[\bar U^2 +
D^2(M\bar M J_{\bar m}) \right].\lb{NU2}
\eea

A simple analysis show that the general solution of these equations has the
form
\be
M = U^2 f, \quad \bar M = \bar U^2 \bar f\,,
\ee
where $f$ is some composite superfunction\footnote{This can be proved, e.g., by
introducing a small parameter before the second terms
in the square brackets in \p{NU2} and representing the solution as a
perturbative series in this parameter. One can show that
each term of this series contains $U^2$ as a factor.}. Thus the integrand
in \p{intmast} contains the nilpotent factor $U^2\bar U^2$ and in the subsequent
manipulations we can use  the effective form of various equations and relations.
For $f$ we obtain in this way the equation
\be
f = -J^{-1}\left( 1 + u f\bar f J_m\right) \quad \mbox{and c.c.}\,,\lb{eqf}
\ee
and also the relation between the variables $u, \bar u$ and $m, \bar m$
\be
m = u \bar f\,, \quad \bar m = \bar u f\,. \lb{mu}
\ee
Substituting this into \p{eqf}, we find the simple representation for $f$
\be
f = - \frac{1}{J + mJ_m}\,, \quad \bar f = - \frac{1}{J + \bar mJ_{\bar m}}\,.
\lb{exprf}
\ee

Now the integrand in \p{intmast} takes the same form as in \p{intaux}, with
\be
E(u, \bar u) = f(u, \bar u) + \bar f (u, \bar u) + f(u, \bar u)
\bar f (u, \bar u)\, J(u, \bar u)\,,
\ee
where $m$ and $\bar m$ are expressed in terms of $u, \bar u$ by Eqs. \p{mu},
\p{exprf}.

If we define
\be
\tilde{E} = u\bar u\,E\,, \quad \tilde{J} = m\bar m\,J\,
\ee
it is straightforward to show that
\be
\tilde{E} = \tilde{J} - m \tilde{J}_m - \bar m \tilde{J}_{\bar m}\,, \quad
 \tilde{J} = \tilde{E}-
u \tilde{E}_u - \bar u \tilde{E}_{\bar u}\,,\lb{Lezh}
\ee
and
\be
u = - \tilde{J}_{\bar m}\,, \; \bar u =  - \tilde{J}_{m}\,, \quad m =
\tilde{E}_{\bar u}\,, \;
\bar m = \tilde{E}_{u}\,.
\ee
These relations are recognized as the basic relations of the $\mu$ representation (Legendre transformation), the only difference
being superfields in place of fields.
Some their useful corollaries directly relating the functions $J$ and $E$ are
\bea
&& J + mJ_m  = -\frac{1}{E + uE_u}\,, \quad \mbox{and c.c.}\,, \lb{JEinv} \\
&& J = -\frac{E + u E_u + \bar u E_{\bar u}}{(E + uE_u)(E + \bar u E_{\bar u})}\,, \quad
E = -\frac{J + m J_m + \bar m J_{\bar m}}{(J + mJ_m)(J + \bar m J_{\bar m})}\,.
\lb{JE}
\eea

We would like to point out once more that all these algebraic relations are valid up to nilpotent
terms vanishing under $U^2\bar U^2\,$.

\subsection{New representation for the ${\cal N}=1$ self-dual systems}
Now we will obtain a new representation for the Lagrangians of nonlinear
electrodynamics in terms of the superfields $W_\alpha$
and $M$, eliminating from the master action \p{mast} the spinor superfield
$U_\alpha$ instead of $M$.

For $U_\alpha$ we obtain the expression
\be
U_\alpha ( 1 + m) = W_\alpha\,, \quad UW = W^2(1 + m)^{-1}\,, \quad
U^2 = W^2 (1 + m)^{-2}\,,
\ee
which, after substitution into \p{mast}, yields the chiral representation for
the action in the $(W,M)$
representation
\bea
&&S(W,M)=\frac14\int d^6\zeta\left[ W^2\,\frac{1-m}{1+m}
-\frac18 \bar{D}^2(M\bar{M}J)\right]+\mbox{c.c.}\,.
\eea
Another form of the same action is
\bea
&&S(W, M) = S_2(W) + S_{int}(W,M)\,, \nn
&&  S_{int}(W,M) = \frac14 \int d^8 z \left[ \left(\frac{1}{1 + m}\,W^2\bar{M}
+ \frac{1}{1 +
\bar m}\,\bar W^2 M\right) + M \bar M\,J\right].\lb{intNW}
\eea

In what follows we will be interested in the self-dual systems, with
$$
J = J(B), \quad B= m\bar m\,.
$$
The equation of motion for $M$ is
\bea
M = -J^{-1}\left[ \frac{1}{(1 + m)^2}\,W^2
+\frac18\bar{D}^2(M \bar M\,\bar{m}J_B)\right].\lb{Nequ}
\eea
Like in the case of Eq. \p{NU2}, the solution of \p{Nequ} has the form
\be
M = W^2\,{\cal B}\,, \lb{NW2}
\ee
where ${\cal B}$ is some composite superfield. Due to the appearance of the
maximal nilpotent factor $W^2\bar W^2$ in \p{intNW}, we can use the fully reduced
effective relations, i.e. make the change ${\cal B} \rightarrow
{\cal B}(w, \bar w)\,$ in \p{NW2} and pass to the
effective equation
\be
\bar{m}={\cal B}(w,\bar{w})\bar{w}.
\ee
Now from \p{Nequ} we obtain
\be
\bar m = - \bar w\,\frac{1}{(1 + m)^2(J + BJ_B)}\,, \quad
m = - w\,\frac{1}{(1 + \bar m)^2(J + BJ_B)}\,.
\lb{mbarm}
\ee
These equations are analogous to the basic equations in the bosonic
$\mu$ representation \cite{IZ1,IZ3},
\be
F^2 = -\bar\mu (1 + \mu)^2\, I_b\,, \quad \bar{F}^2=-\mu(1+\bar\mu)^2\,I_b\,,
\; b = \mu\bar\mu\,,
\ee
with the obvious correspondence
\be
(w, \bar w)\;\leftrightarrow \;(\bar F^2, F^2)\,, \quad (m, \bar m) \;
\leftrightarrow \; (\mu, \bar\mu)\,,
\quad (J + BJ_B) \;\leftrightarrow \;I_b\,.
\ee

Solving  Eqs. \p{mbarm} for $m$ and $\bar m$ in terms of $w, \bar w$ and
substituting the solution into \p{intNW}, we can find the relevant
self-dual ${\cal N}=1$ action in terms of the superfield strengths
$W_\alpha, \bar W_{\dot\alpha}$.
Like in the bosonic case, only for some special
superfunctions $J$ these equations have the solution in a closed form,
while in other cases one manages to obtain the action only as a power series
in $w, \bar w$. Nevertheless, the way to the final $W, \bar W$ action in
the $(W,M)$ representation in some cases turns out to be easier than in the
original $(W, U)$ representation, which deals with the variables
$u, \bar u$ instead of $m, \bar m\,$. Despite this technical difference,
both representations (at least for the considered particular set of
self-dual systems, with $\Lambda = \Lambda(w, \bar w)$) are equivalent
to each other. The basic objects in both representations are related to
each other by the relations \p{mu}, \p{exprf}, \p{JEinv} and \p{JE}
specialized to the $U(1)$ invariant case, e.g.,
 \bea
\frac{d}{dA}(AE)=-\left[\frac{d}{dB}(BJ)\right]^{-1},\quad A=B\left[\frac{d}{dB}(BJ)\right]^2,
\quad B=A\left[\frac{d}{dA}(AE)\right]^2\,,
\eea
with  $A = u\bar u\,, \,B = m\bar m\,$.

\setcounter{equation}0
\section{Examples of the ${\cal N}=1$  self-dual models}

\subsection{${\cal N}=1$ Born-Infeld}
The superfield action for the ${\cal N}{=}1$ BI theory  can be rewritten
in the polynomial form by making use of the auxiliary complex superfields
$X$ and $R$
\bea
&&S(W,X,R)=S_2(W,\bar{W})+S_{int}(W,X,R),\\
&&S_{int}(W,X,R)=\frac18\int d^8z\Big\{X\bar{X}
-\bar{R}\Big[X+\frac1{16}\bar{D}^2(X\bar{X})-W^2\Big]\nn
&&
-R\Big[\bar{X}+\frac1{16}D^2(X\bar{X})-\bar{W}^2\Big]\Big\}.\lb{BIpolyn}
\eea
Varying it with respect to  $R\,$, we obtain the constraint
\be
W^2=X+\frac1{16}\bar{D}^2(X\bar{X})\sim X+\frac1{16}X\bar{D}^2\bar{X}\,,
\lb{BIgen}
\ee
which guarantees chirality of the  solution, $\bar{D}_{\dot\alpha} X=0$.
Note that this constraint yields the dimensionless effective relation
\be
w=x+\frac12x\bar{x}, \quad x=\frac18\bar{D}^2\bar{X},\quad \bar{x}=\frac18D^2X\,,
\ee
which is equivalent to the algebraic equation in the bosonic BI theory.

The superfield action \p{BIpolyn} becomes
\bea
S(W,X)=S_2(W,\bar{W})+\frac{1}{8}\int d^8zX\bar{X}\,,
\eea
which can be shown to be equivalent to \p{BIX}.

On the other hand, the superfield $X$ can be eliminated via its equation
of motion
\bea
&&X-R-\frac1{16}X\bar{D}^2\bar{R}-\frac1{16}XD^2R=
X-R-\frac1{2}X(r+\bar{r})=0\,, \lb{XR1}
\eea
where $r=\frac18\bar{D}^2\bar{R}\,, \quad \bar{r}=\frac18D^2R\,$.
Now we wish to solve this equation for
$X$ in terms of $R, \bar R$ and to finally get the
$R, W$ form of the action \p{BIpolyn}. The exact solution of \p{XR1} is
as follows
\bea
X=\frac{R}{(1-\frac12r-\frac12\bar{r})}\,.\lb{XR}
\eea

Substituting this solution for $X$ in the action \p{BIpolyn} we obtain
\bea
S_{int}(W,R)=\frac18\int d^8z\left[(\bar{R}W^2+R\bar{W}^2)
-\frac{R\bar{R}}{(1-\frac12r-\frac12\bar{r})}\right].\lb{SWR}
\eea
This action is recognized as a particular case of our representation \p{intNW}
after redefining the auxiliary variables as
\bea
\frac{2M}{1+\bar{m}}=R\,, \quad
\bar{r}=\frac{2\bar{m}}{1+\bar{m}}\,.
\eea
The inverse relation involves the differential operator $V$
\be
M=\frac1{2(1-V)}R\,,\quad V:=\frac1{16}RD^2,\quad \bar{m}=
\frac{\bar{r}}{2-\bar{r}}\,.
\ee
The transformation law for the superfield $R$ follows from the
$U(1)$ transformation \p{Ntrans} of $M$ and $\bar{m}=\frac18D^2M$:
\bea
\delta R=i\omega R(2-\bar{r})\,,\quad \delta \bar{r}=i\omega \bar{r}(2-\bar{r})\,.
\eea

The action \p{SWR} yields the auxiliary equation
\bea
W^2-\frac{R}{1-\frac12r-\frac12\bar{r}}-\bar{D}^2
\left[\frac{R\bar{R}}{16(1-\frac12r-\frac12\bar{r})^2}\right]=0\,,
\eea
which is a particular case of \p{Nequ}. Using \p{XR}, we can bring this
relation to
the form of the constraint \p{BIgen}.

The ${\cal N}{=}1$ BI model in our $(W,m)$ formalism
corresponds to the choice of the invariant density
\bea
&&J^{\B\I}=\frac{2}{B-1}\,,\quad J^{\B\I}+BJ^{\B\I}_B=-\frac{2}{(B-1)^2}\,,\\
&&\bar{w}=\frac{2\bar{m}(1+m)^2}{(m\bar{m}-1)^2}\,. \lb{mBI}
\eea
The effective equation for $\bar{w}$ is similar to the bosonic relation
for $F^2$
in the $\mu$ representation of the $BI$ theory \cite{IZ1,IZ3}.

Eq. \p{mBI} and its conjugate can be solved for $m, \bar m$ in terms of
$w, \bar w$, which finally
reproduces the ${\cal N}=1$ BI action \p{N1BI}
\bea
&&\bar{m}=\frac{Q-1+\frac12(w-\bar{w})}{Q+1-\frac12(w-\bar{w})}\,,\\
&&Q(w,\bar{w}) = \sqrt{1+w+\bar{w}+(1/4)(w-\bar{w})^2}\,,\\
&&G=1-w\Lambda-w\bar{w}\Lambda_{\bar{w}}
=\frac1{1+\bar{m}}=\frac1{2Q}\left[Q+1-\frac12(w-\bar{w})\right].
\eea

The ${\cal N}=1$ BI theory in the original $(W,U)$ representation corresponds to
the choice
\bea
&&{\cal P}(A)=\frac{d}{dA}(AE^{\B\I})=\frac12(B-1)^2,\quad A =
\frac{4B}{(1-B)^4}\,,\\
&&2{\cal P}=[1-A{\cal P}^2]^2\,,
\eea
whence $E^{\B\I}(A)=\frac12-\frac18A+\frac3{32}A^2+O(A^3)\,$.

\subsection{Other examples}
The ${\cal N}{=}1$ analog of the bosonic simplest interaction model
\cite{IZ,IZ3} corresponds to the
choice $J=-2$ or $E=\frac12$. The corresponding algebraic equations read
\bea
&&W_\alpha=U_\alpha+\frac{1}{16}U_\alpha
\bar{D}^2\bar{U}^2=U_\alpha\left(1+\frac12 u\right),\lb{1paramet}\\
&&W^2=U^2\left(1+\frac12 u\right)^2,\quad \bar{w}|=\bar{u}\left(1+\frac12 u\right)^2.
\eea
The perturbative solution for $u(w,\bar{w})$ is completely similar
to the corresponding bosonic solution in the simplest interaction model
\cite{IZ3}
\bea
\Lambda_{\s SI}= \frac12 - \frac14(w + \bar w) + \frac12w\bar w +
\frac18 (w^2 + \bar w^2)\\
- \frac{5}{8}w\bar w(w + \bar w)
 - \frac1{16}(w^3 + \bar w^3) + O(W^8)\,.\lb{SIsuper}
\eea

By analogy with bosonic invariant interaction $I_b=\frac{2}{b-1}$
\cite{IZ3} we can obtain the exact formula for $\Lambda(w,\bar{w})$
for the choice
\be
J+BJ_B=\frac{2}{B-1}\,.
\ee
In this case the effective equation\p{mbarm} is reduced to the
solvable cubic equation for the real superfield $\hat{r}$
\bea
m=\hat{r}-\frac14\,(w-\bar{w})\,.
\eea

We can also study the simple example of the invariant interaction with the
additional variable $g$
\be
E=\frac12+f_1\,g\bar{g}\,,
\ee
where $f_1$ is a real constant. The basic recursive auxiliary equation in
this case is
\bea
U_\alpha=W_\alpha
-\frac1{16}U_\alpha\bar{D}^2\,\bar{U}^2
+\frac{f_1}{8}\,U_\beta\bar{D}^2\,(\bar{U}^2D_\alpha U^\beta \bar{g})\,.
\lb{gequ}
\eea
The 5-th order term in its perturbative solution has the form
\bea
&&U^{(5)}_\alpha=\frac12W_\alpha (w\bar{w}+\frac12\,w^2)
+f_1\,w\bar{y}\,W_\beta D_\alpha W^{\beta}.
\eea
Using  Eq.\p{UWW}, we obtain the corresponding term in $\Lambda$:
\bea
\Lambda^{(2)}_y=f_1\bar{y},\quad \Lambda^{(2)}=- \frac14(w + \bar w)+f_1y\bar{y}\,.
\eea
Thereby we restore the nonlinear superfield action up to the 6-th order in
$W_\alpha, \bar W_{\dot\alpha}\,$.

\subsection{ ${\cal N}=1$ self-dual models with higher derivatives}
In supersymmetric self-dual models with higher derivatives
we still can use  our basic bilinear action
$S_2(W,U)$ \p{freeWV}, which corresponds to the usual free equations without
additional derivatives. This choice guarantees the $U(1)$ duality of the
entire equations of motion, if we construct  the $U(1)$ invariant
interaction with higher derivatives
solely in terms of auxiliary superfields.  In the $W$ representation the
action will involve powers
of some basic coupling
constant $c$ of dimension $-2$, as well as plenty of additional dimensionless
coupling constants.

Thus the self-dual interactions with higher derivatives can be naturally
introduced via
the modification of the auxiliary invariant interaction.
The possible bilinear interaction
\be
ca_1\int d^8z\bar{U}^{\dot\alpha}\partial_{\alpha\dot\alpha}U^\alpha\,,\lb{Uprop}
\ee
drastically changes the status of the superfield $U_\alpha$
which will become propagating.
In components, this would mean, in particular, that the scalar component field
$v(x)$ of $U_\alpha$ (and, respectively,  $D(x)$ in the $W$ representation) will propagate. This effect disappears in the limit
$a_1\rightarrow 0$. Such self-dual deformations of the bilinear bosonic action were considered, e.g., in \cite{BN,CKO},
and the interactions of the type \p{Uprop} would give ${\cal N}=1$ self-dual superextensions of these deformed actions.
Further in this Subsection we will focus  on the deformations which do not change the free action.

As an example, let us consider the $U(1)$ invariant quartic interaction of the
auxiliary superfields with higher derivatives
\bea
&&\frac14c\int d^8z\,b_1\, \partial^mU^2\partial_m\bar U^2\lb{2der}\,,
\eea
where $b_1$ is a dimensionless  coupling constant.
The  auxiliary equation has the form
\bea
&&W_\alpha=U_\alpha+\frac18cb_1U_\alpha\Box\bar{D}^2\bar{U}^2 =
U_\alpha\left(1+cb_1\,\Box u\right),\\
&&\bar{w}|=\bar{u}\left(1+cb_1\Box u \right)^2\,.
\eea
Its perturbative solution is as follows
\bea
&&U_\alpha|=W_\alpha\left[1-cb_1\Box w+c^2b^2_1(\Box w)^2+2c^2b^2_1\Box(w\Box \bar{w})
+\ldots\right]\,,
\eea
where $\Box=\partial^m\partial_m$. Using this solution, we can construct
the corresponding self-dual nonlinear action with higher derivatives in the $W$
representation.

We can also study an example of the 6-th order invariant interaction
with higher derivatives
\bea
&&\frac14cb_2\int d^8zU^2\bar{U}^2\partial_mg\partial^m\bar{g}\lb{patg}
\eea
and find the power-series solution of the corresponding auxiliary equation.

Further generalizations involve $U(1)$ invariant interactions with derivatives multiplied by
some polynomials in the dimensionless variables $u,\; \bar{u},\; g,\; \bar{g}\,$.

Using the invariant interactions of the type
\bea
&&\frac14c^2\int d^8z \,b_4(\Box U^2)(\Box \bar{U}^2)\,,
\eea
we arrive at the self-dual theories with the growing powers
of higher derivatives. In this case, the auxiliary equation has the form
\bea
&&W_\alpha=U_\alpha+\frac18c^2b_4U_\alpha\Box^2\bar{D}^2\bar{U}^2=U_\alpha
(1+c^2b_4\Box^2 u)\,.
\eea

To summarize, the principle that the higher-derivative actions in the
${\cal N}{=}1$ electrodynamics models are generated by some
higher-derivative $U(1)$ invariant interactions of the auxiliary spinor
superfields automatically
yields the self-dual nonlinear actions
in the $W$ representation.

\setcounter{equation}0
\section{Bosonic limit}
In this Section we consider the bosonic component Lagrangians corresponding to some
superfield ones considered
above. Our conventions on
the bosonic component fields are
\bea
&&W_\alpha = 2 iF_\alpha^{\;\;\beta}\theta_\beta - \theta_\alpha D +
\frac i2 \theta^2
(\sigma^m\bar\theta)_\beta(\delta_\alpha^\beta \partial_m D
-2i \partial_m F_\alpha^{\;\;\beta})\,, \quad
F_{\alpha}^{\;\;\beta} = \frac{1}{4}(\sigma^m\bar\sigma^n)_\alpha^\beta F_{mn}\,, \nn
&&U_\alpha = 2i V_\alpha^{\;\;\beta}\theta_\beta - \theta_\alpha v +
\frac i2 \theta^2(\sigma^m\bar\theta)_\beta(\delta_\alpha^\beta \partial_m v
- 2i\partial_m V_\alpha^{\;\;\beta})\,.\lb{boson}
\eea
Here, the symmetric complex bispinor field $V_{\alpha\beta}(x) =
V_{\beta\alpha}(x)$  and the complex
scalar field $v(x)$ are not subject
to any constraints off shell. We also have $F_{mn} = \partial_m A_n -
\partial_n A_m$ and $D = \bar D$
in virtue of the superfield Bianchi identity
\p{Bianchi}.

Various superfield objects constructed from $W_\alpha$ and $U_\alpha$ have
the following bosonic limits
\bea
&&W^2 \,\rightarrow\, (-2\varphi + D^2)\theta^2\,, \; U^2 \,\rightarrow\,
(-2\nu + v^2)\theta^2\,, \;
UW   \,\rightarrow\, (-2VF + vD)\theta^2\,, \nn
&& w \,\rightarrow\, (\bar\varphi - \frac12 D^2)\,, \; y, \bar y\,\rightarrow\,
2D\,, \;
u \,\rightarrow\, (\bar\nu - \frac12 \bar v^2)\,,
\; g \,\rightarrow\, 2v\,, \lb{2line}
\eea
where
$$\varphi = F^{\alpha\beta}F_{\alpha\beta} =  \frac14 F^{mn}F_{mn}
+\frac{i}4
F^{mn}\tilde{F}_{mn}\,, \quad\nu = V^{\alpha\beta}V_{\alpha\beta}$$ as
in \cite{IZ3}. In the second line of \p{2line} we took into account that the relevant
superfield arguments appear
in the superfield actions always with the
nilpotent factors $W^2\bar W^2$ or $U^2\bar U^2$ and so, after integration
over $\theta, \bar\theta$
and taking the bosonic limit, are
reduced to their lowest $\theta = \bar\theta =0$ components.
The study of the component bosonic equations can give us a further insight into the
properties of the superfield equations for $U^2(W)$  and $D^\alpha U_\alpha(W)$
\p{wueff}, \p{DWeff}.

The free actions \p{free} and \p{freeWV} are reduced to
\bea
&&S_2(W) \; \rightarrow \; \int d^4 x \left[ -\frac12(\varphi + \bar\varphi)
+ \frac12 D^2\right] =
\int d^4 x \left( -\frac14 F^{mn}F_{mn}  + \frac12 D^2\right), \lb{compW2} \\
&& S_2(W,U) \; \rightarrow \; \int d^4 x \left(-2 VF + \nu +\frac12\varphi +
vD-\frac12 v^2 -\frac14 D^2 + \mbox{c.c.}  \right).\lb{compWU2}
\eea
After integrating out the auxiliary fields $V_{\alpha\beta}, v$ from
\p{compWU2} we recover \p{compW2}.

The interaction
 \p{intaux} is reduced to
\bea
&&S_{int}(W, U)\; \Rightarrow \;\int d^4x\, (\nu - \frac12 v^2)(\bar\nu -
\frac12 \bar v^2)
E[ (\nu - \frac12 v^2), (\bar\nu - \frac12 \bar v^2), 2v, 2\bar v]\,.\lb{compWU}
\eea
Now it becomes clear why using of the alternative superfield variables
\p{prime}, \p{prime2}
looks more preferable: these quantities have a simpler bosonic limit
\bea
&&u' \,\rightarrow\, \bar\nu\,, \quad \bar u' \,\rightarrow\, \nu\,, \\
&&E \; \rightarrow \; E'(\nu, \bar\nu, v,\bar v)\,,\quad \Lambda
\; \rightarrow \;\Lambda'(\varphi,\bar\varphi,D)\,.
\eea
Correspondingly, for the invariant interactions we have
\bea
&&A' \,\rightarrow\,
a := \nu\bar\nu\,, \quad
B'\,\rightarrow\,\bar\nu v^2\,, \quad
\bar B'\,\rightarrow\, \nu \bar v^2\,,\\
&&{\cal F}(A,B,C)\,\rightarrow\, {\cal F}'(\nu\bar\nu,v^2\bar\nu,v\bar{v})\,.
\eea

Just this choice of $E$ is most convenient for examining the role of the scalar
auxiliary fields $v, \bar v, D\,$.
For these fields we obtain the following equations of motion:
\bea
&&\delta D: \quad D = v + \bar v\,, \lb{DV}\\
&&\delta v: \quad D-v -  v (\bar\nu - \frac12 \bar v^2) E' +
(\nu - \frac12 v^2)(\bar\nu - \frac12 \bar v^2)
E'_v = 0\,. \quad
\lb{vbarv}
\eea
For the self-dual case $E'_v$
is proportional to $v$ or $\bar v$,
so Eq. \p{vbarv} and its conjugate, after eliminating $D$ by Eq. \p{DV},  are reduced to a system of two homogeneous equations
for $v, \bar v\,$, such that the determinant of the $2\times 2$ matrix of the coefficients is non-vanishing at the origin.
This means that the main perturbative
 solution of
\p{vbarv} in the duality-invariant case is
\be
v = \bar v = 0 \; \Rightarrow \;  D=0\,.\lb{triv}
\ee
To make sure, we  analyzed the nonlinear equation \p{vbarv}
for the simple interaction  $E'=e_1+f_1\,g\bar{g}\, \Rightarrow \,e_1+ 4 f_1\,v\bar{v} $ and did not find
any nontrivial analytic solution $v \neq 0\,$. Nevertheless, the existence of some non-perturbative non-trivial
solutions for $v, \bar v$ (and $D$) under some special choices of $E$ (or $E'$) cannot  be excluded.

For the basic solution \p{triv}, $S_{int}(W, U)$ in the bosonic limit  becomes
\bea
S_{int}(W, U)\; \rightarrow \;\int d^4x\, a\, E (a)\,.
\eea
Comparing it with the general auxiliary self-interaction in the tensorial
auxiliary field formulation of
the bosonic self-dual Maxwell models \cite{IZ2,IZ3}, we identify
\be
a E(a) = {\cal E}(a)\,.
\ee

Interactions with higher derivatives change Eq. \p{vbarv}.
In particular,  the superfield interaction  \p{patg}
 yields  the component term
\bea
4cb_2\int d^4x (\nu - \frac12 v^2) (\bar\nu - \frac12 \bar v^2)
\partial^mv\partial_m\bar{v}
\eea
in addition to the bosonic action \p{compWU2}.
Once again, solving recursively the equations for $v, \bar v$, we find the
trivial perturbative solution \p{triv}
as the unique one.

The Fayet-Iliopoulos (FI) term $\xi D$ softly breaks  the $U(1)$ duality and
deforms the $\delta D$ equation to $\xi+D=v+\bar{v}\,$. The models with this
term added provide  nontrivial solutions
for the auxiliary fields $v$ and $D$, depending on the parametrization of $E'$
\footnote{ A nontrivial FI-term deformed solution for the field  $D$ was
considered in  \cite{Ku1} for the case of  ${\cal N}{=}1$ BI theory.}.

The bilinear invariant interaction \p{Uprop} gives the bosonic Lagrangian
$\sim ca_1\partial^mv\partial_m\bar{v}$ and so radically affects  the
component equation \p{vbarv}, yet preserving self-duality. The former auxiliary
fields $v, \bar v$ become propagating in this case,
while elimination of $D$ produces ``mass terms'' for these fields.

It is natural to  treat both this bilinear interaction and the FI term as a
kind of  non-perturbative effects generating nontrivial solutions for the
auxiliary fields. So in the presence of such terms the dependence of the
auxiliary interaction $E$ on the additional superfield variables
$(g, \bar g) \sim (v, \bar v)$ can prove very essential.

\setcounter{equation}0
\section{$U(N)$ duality for ${\cal N}=1$}
\subsection{Auxiliary chiral $U(N)$ superfields}
Let us consider $N$  Abelian superfield strengths
\bea
W_\alpha^i=-{1\over4}\bar{D}^2A_\alpha^i=-{1\over4}\bar{D}^2D_\alpha V^i ~,\quad
\bar W_{\dot\alpha}^i=-{1\over4}D^2\bar
A_{\dot\alpha}^i= -{1\over4}D^2\bar{D}_{\dot\alpha} V^i\,, \; i = 1, \ldots N\,,
\eea
with $V^i$ being $N$ real gauge prepotentials.
By definition, all superfields are transformed by the vector representation of
the group $O(N)$:
\be \delta
W^i_\alpha=\xi^{ik}W^k_\alpha~,\quad \delta\bar{W}^i_{\dot\alpha}=
\xi^{ik}\bar{W}^k_{\dot\alpha}\,,
\ee
where
$\xi^{ik}=-\xi^{ki}$ are real group parameters. The free action of this set of
superfields is
\be
S_2(W^i,\bar W^i)=\frac1{4}\int d^6\zeta
(W^iW^i)+\mbox{c.c.}\,, \lb{Nbil}
\ee
where $(W^iW^i) :=W^{i\alpha}W^i_\alpha\,$.

Further, we introduce the notation
\bea
W^{kl}=(W^{k\alpha}W^l_\alpha),\quad\bar{W}^{kl}=(\bar{W}^k_{\dot\alpha}
\bar{W}^{l\dot\alpha}) \lb{defWkl}
\eea
and consider the particular parametrization of the nonlinear $O(N)$ and
$R$ invariant superfield interaction
\bea
&&S_\Lambda(W^{kl},\bar W^{kl})=
\frac1{4}\int d^8z W^{kl}\bar{W}^{rs}\Lambda_{kl,rs}(w,\bar{w})\,,
\eea
where $\Lambda_{kl,rs}(w,\bar{w})$ is a function of the dimensionless
Lorentz invariant matrix variables
\bea
&&w^{kl}=\frac18\bar{D}^2\bar{W}^{kl},\quad
\bar{w}^{kl}=\frac18D^2W^{kl}\,.
\eea

The integral conditions of $U(N)$ duality have the following form:
\bea
&&\mbox{Im}\int d^6\zeta[(W^iM^k)-(W^kM^i)]=0~,\lb{ON} \\
&&\mbox{Im}\int d^6\zeta[(W^iW^k)+(M^iM^k)]=0~,\lb{nonlU}
\eea
where
\bea
M_\alpha^k\equiv-2i\frac{\delta S}{\delta W^{k\alpha}}~, \lb{defMN}
\eea
and $(W^iW^k), (W^iM^k)$ and $(M^iM^k)$ are defined similarly to \p{defWkl}.
The antisymmetric in $i$ and $k$ condition \p{ON} means the off-shell
$O(N)$ symmetry, so the nontrivial constraint in the general case is
the nonlinear condition \p{nonlU}.

Consider the following transformations
\be
\delta_\eta
W^i_\alpha=\eta^{ik}M^k_\alpha~,\quad \delta_\eta M^i_\alpha=
-\eta^{ik}W^k_\alpha~, \lb{UNON}
\ee
where $\eta^{ik}=\eta^{ki}$ are
real parameters. The complex combination
\be
U^k_\alpha={1\over2}(W^k_\alpha+i M^k_\alpha) \lb{defUN}
\ee
transform linearly in the group $U(N)$ according to the fundamental
representation of the latter:
\be
\delta
U^k_\alpha=(\xi^{kl}-i\eta^{kl})U^l_\alpha\,.
\ee
The covariance of the set of Bianchi identities for $W_\alpha^i$ together
with the superfield equations of motion
following from the action $S_2 + S_\Lambda$ under the coset $U(N)/O(N)$
transformations \p{UNON} is the correct generalization
of the notion of $U(1)$ self-duality to the considered case.  The condition
\p{nonlU} ensures compatibility of this duality
covariance with the definition \p{defMN}.

The basic steps in generalizing the $U(1)$ self-duality setting with the
single auxiliary superfield $U_\alpha$  to the $U(N)$ case is to
interpret $U^k_\alpha$ defined in \p{defUN}  as auxiliary chiral
superfields $(R(U^i_\alpha)=1)$ and
to replace \p{Nbil} by the following bilinear action:
\be
S_2(W^k,U^k)=\int
d^6\zeta[(W^kU^k)-\frac14(W^kW^k)-\frac12(U^kU^k)]
+\mbox{c.c.}\,.\lb{WUN2}
\ee
The corresponding auxiliary interaction $S_E(U)$ is chosen as an
arbitrary $O(N)$ and R invariant functional  of
the auxiliary superfields $U^k$ and $\bar{U}^k$ (and perhaps of
their derivatives). The basic equation for the
superfield $U^k_\alpha$ is
\be
W_\alpha^k=U_\alpha^k-\frac{\delta S_E(U)}{\delta U^{k\alpha}}\,.\lb{WUNeq}
\ee
Like in the $U(1)$ case, it is straightforward to show that the $U(N)$
duality conditions \p{nonlU}
amount to the $U(N)$-invariance of $S_E(U)\,$:
\be
\int d^6\zeta U^k_\alpha\frac{\delta
S_E(U)}{\delta U^l_\alpha}- \int d^6\bar\zeta\bar
U^k_{\dot\alpha}\frac{\delta S_E(U)}{\delta \bar{U}^l_{\dot\alpha}}=0\,.
\ee

A particular parametrization of $S_E$ is through independent dimensionless
R-invariant Lorentz scalars
\bea
&&u^{kl}={1\over8}\bar{D}^2\bar U^{kl}~,\quad \bar{u}^{kl}=
{1\over8}D^2 U^{kl}\,,\lb{ukl} \\
&&\delta_\eta u^{kl}=-i\eta^{kr}u^{rl}-i\eta^{lr}u^{rk},\quad
\delta_\eta \bar{u}^{kl}=i\eta^{kr}\bar{u}^{rl}+i\eta^{lr}\bar{u}^{rk}
\eea
(these are analogs of the variables $u$ and $\bar u$ of the $U(1)$ case).
Then in the self-dual theory
we can consider the following particular $R$- and $U(N)$ invariant interaction
of the auxiliary superfields
\be
S_E=\frac14\int d^8z(U^{l}U^k)(\bar{U}^r\bar{U}^{s})E_{kl,rs}\,,
\ee
where $E_{kl,rs}$ is
 the $U(N)$ covariant dimensionless superfield density composed out of the
 variables \p{ukl}.

A simple example of the action functional of this type contains the matrix
$E^{lk}(A)$, which depends on the matrix argument $A^{kl}=\bar{u}^{kr}u^{rl}$:
\be
S_E=\frac14\int d^8z(\bar{U}^k\bar{U}^{s})(U^{s}U^l)E^{lk}(A)\,.\lb{Ematr}
\ee

We can also consider the interaction with a scalar invariant density,
\bea
S_E=\frac14\int d^8z (U^kU^l)(\bar{U}^l\bar{U}^k)E(A_n)\,,\lb{Escal}
\eea
where the  dimensionless invariant variables $A_n$ are defined as follows
\bea
&&
 A_n=\frac1n\,\mbox{Tr}\,A^n\,.\lb{invvar}
\eea
Using the relations
\bea
\delta A_n=\frac14D^2(\delta U^k  U^r)(uA^{n-1})^{rk}\,,\quad
(uA^{n-1})^{rk}=(uA^{n-1})^{kr}\,,
\eea
we derive the  equation of motion for the auxiliary spinor superfield
in this case
\bea
&&W^k_\alpha-U^k_\alpha
=\frac1{8}U^l_\alpha
\bar{D}^2\Big\{ (\bar{U}^l\bar{U}^k)E(A_n)+\frac18(\bar{U}^p\bar{U}^t)
D^2[(U^tU^p)E_nu^{ls}(A^{n-1})^{sk}]\Big\}\,,
\eea
where $E_n=\partial E/\partial A_n\,$. This $U(N)$ covariant superfield
equation describes the particular class of  self-dual models.

\subsection{$U(N)$ analog of the $M$ representation}
An alternative representation for the supersymmetric $U(N)$ self-dual
theories deals with  $W^k_\alpha$, $U^k_\alpha$ and, in addition, with
the auxiliary general scalar superfields $M^{kl}=M^{lk}$ and their
dimensionless derivatives  $\bar{m}^{kl}=\frac18D^2M^{kl}$ (as well as
with the corresponding conjugated superfields).
 Under the duality group $U(N)$ the new auxiliary superfields are
 transformed as
\bea
&&\delta_\eta\bar{M}^{kl}=i\eta^{kr}\bar{M}^{rl}+i\eta^{lr}\bar{M}^{kr}\,,\quad
\delta M^{kl}=-i\eta^{kr}M^{rl}-i\eta^{lr}M^{kr}\,,
\\
&&\delta_\eta m^{kl}=i\eta^{kr}m^{rl}+i\eta^{lr}m^{kr},\quad
\delta_\eta \bar{m}^{kl}=-i\eta^{kr}\bar{m}^{rl}-i\eta^{lr}\bar{m}^{kr}\,.
\eea

The general ``master'' action is a sum of  the bilinear action $S_2(W,U)$
\p{WUN2} and the $U(N)$ invariant interaction
\bea
&&S_{int}( U, M) = \frac14 \int d^8z[(U^kU^l)\bar{M}^{kl} +
(\bar{U}^k\bar{U}^l) M^{kl}] +S_{int}(M)\,,\\
&&S_{int}(M)=\frac14 \int d^8z
\bar{M}^{kl}M^{rs}\,J_{kl,rs}\,(m, \bar m) \,,
\lb{Nintmast}
\eea
where $J_{kl,rs}$ is a dimensionless covariant density. So, the master
action is
\be
S(W,U,M) = S_2(W,U) + S_{int}( U, M)\,.\lb{SWUM}
\ee
For the density $J_{kl,rs}$ we can choose, e.g.,  the following particular
parametrization:
\bea
&&J_{kl,rs}=\frac14(\delta^{ks}J^{rl}+\delta^{ls}J^{rk}+\delta^{kr}J^{sl}
+\delta^{lr}J^{sk})\,,\lb{JBmatr}\\
&&\delta_\eta J^{lk}(B)=i\eta^{ls}J^{sk}-iJ^{ls}\eta^{sk}\,,\nonumber
\eea
where $J^{lk}(B)$ is a matrix function of $B^{ij}=m^{is}\bar{m}^{sj}\,$,
for instance,
\be
J^{lk}=-2\delta^{lk}+\frac12i_2B^{lk}+\ldots\,.\lb{JBNpert}
\ee

 Varying \p{SWUM} with respect to $U^k_\alpha\,$, we obtain
the equation
\be
W^k_\alpha=U^l_\alpha(\delta^{kl}+m^{kl})\,.\lb{WUNeq1}
\ee
Then, eliminating the variables $U^k_\alpha\,$, we come to the $U(N)$ analog of
the representation \p{intNW}
\bea
&&S(W^{kl}, M^{kl})=S_2(W,\bar{W})+S_{int}(M)\nn
&&+\frac14\int d^8z\mbox{Tr}\left[W\left(\frac{\bf 1}{{\bf 1}+m}\right)\bar{M}
+\bar{W}\left(\frac{\bf 1}{{\bf 1}+\bar{m}}\right)M\right]\lb{SWMN}
\eea
where ${\bf 1}$ denotes the unit matrix.
Varying \p{SWUM} with respect to the superfields $\bar{M}^{kl}$, we obtain
the $U(N)$ analog of the
equations \p{Nequ}. This equation can be solved perturbatively. The matrix
solution $M(W,\bar{W})$ yields the self-dual superfield action in
the $W$ representation.

The parametrization of $J_{kl,rs}$, which is yet simpler than \p{JBmatr},
involves only one invariant function $J$
\bea
J_{kl,rs}=\frac12(\delta_{kr}\delta_{ls}+\delta_{ks}\delta_{lr})J(m,\bar{m}),
\quad S_{int}(M) =\frac14 \int d^8z \bar{M}^{kr}M^{rk}J(B_n)\,.\lb{JBNscal}
\eea
The variables $B_n$ on which the invariant function $J$ depends are defined as
\be
B_n=\frac1n\,\mbox{Tr}B^n\,.
\ee
The simplest possible interaction is $J(B_1)$, and it corresponds to the
special choice of the invariant density in \p{Escal} as
$E(A_1)\,$.

\subsection{Examples of the $U(N)$ self-dual theories}
As an example of the $U(N)$ self-dual action, we may consider the
following simplest interaction:
\be
S_{SI}=\frac18\int
d^8z\,(U^kU^l)(\bar{U}^k\bar{U}^l)\,,
\ee
which gives us the basic algebraic
spinor equations in the form
\bea
&&W_\alpha^k=U_\alpha^k+\frac1{16}U_\alpha^l\bar{D}^2(\bar{U}^k\bar{U}^l)\,.
\eea
Using the perturbative solution $U_\alpha^k(W,\bar{W})$, we obtain
an $U(N)$ analog of our model \p{SIsuper}
\bea
&&S_\Lambda=\frac18\int d^8z\mbox{Tr}\Big[W\bar{W}-\frac12(W\bar{w}\bar{W}
+Ww\bar{W})+O(W^4)\Big].
\eea

 The $U(N)$ supersymmetric generalization of the BI model is based
on the following
 matrix algebraic relation \cite{ABMZ,KT}
\bea
&&X^{kl}+\frac1{16}X^{kj}\bar{D}^2\bar X^{lj}
=W^{kl}\,,\lb{XNeq}
\eea
where $X^{kl}\neq X^{lk}$ are the auxiliary chiral superfields,
and $\bar{X}^{jl}$ are the conjugated antichiral superfields.
The nonsymmetric matrix $W^{kl}$ corresponds to the so-called $U(N)\times U(N)$
duality.

In our formulation this model corresponds to the $U(N)$ invariant
representation \p{JBmatr} with
\be
J^{lk}=2\left(\frac{\bf 1}{B-{\bf 1}}\right)^{lk}\,.
\ee
The equivalence of the two formulations can be checked  by comparing
the perturbative expansions of the relevant actions $S_\Lambda(W,\bar{W})$ .

In our formalism we can also consider alternative $U(N)$ self-dual
generalizations of the supersymmetric $U(1)$ BI model. For instance,
we can use \p{JBNscal} with the one-parameter invariant interaction
\be
J(B_1)=\frac2{B_1-1}\,, \quad B_1=m^{kl}\bar{m}^{lk}\,.
\ee

\setcounter{equation}0
\section{Conclusions}
In this paper, we constructed the most general ${\cal N}{=}1$ superextension
of the
auxiliary bispinor field formulation of the $U(1)$ duality-invariant nonlinear
electrodynamics models which was proposed in \cite{IZ1,IZ3}.
The auxiliary bispinor fields are accommodated by the auxiliary spinor
superfield
and the full set of self-dual ${\cal N}{=}1$ models is
parametrized by $U(1)$ duality-invariant self-interactions of this superfield.
The conventional nonlinear action in terms of
the Maxwell superfield strengths is reproduced as a result of elimination of
the auxiliary superfield
by its equation of
motion.

As compared with the recent paper \cite{Ku} devoted to the same issue of
${\cal N}{=}1$ supersymmetrizing of the formulation with bispinor fields,
we allow for the most general dependence of the auxiliary superfield Lagrangians
on the $U(1)$ duality-invariant superfield arguments.
Though the dependence on the extra $U(1)$ invariant superfield variables
gets seemingly inessential on shell,  when considering
the ``pure'' ${\cal N}{=}1$ self-dual systems which deal with the Maxwell
superfield strengths only, it can become essential and capable to
provide new models in the cases of various deformations of such systems, e.g.,
through adding the Fayet-Iliopoulos term to the action
\cite{Ku1} or turning on the couplings to the charged chiral matter.  Other new
results of our study is the construction of ${\cal N}{=}1$ generalization of
the so called $\mu$ version of the approach of \cite{IZ1,IZ3} (which
significantly
simplify various computations), and finding out how to generate self-dual
${\cal N}{=}1$ systems
with higher derivatives from
the appropriate modifications of the $U(1)$ invariant auxiliary interaction.
In the latter case the extra superfield variables
we have introduced can play an essential role: they indeed considerably enlarge
the set of possible $U(1)$ invariant interactions, and these additional
interactions are not trivialized
on shell in some conceivable cases.

We also presented a few explicit examples of generating duality-invariant
${\cal N}{=}1$ superfield systems in the approach with auxiliary
spinor superfield, as well as gave the bosonic component Lagrangians for the
general case, with the auxiliary fields being kept,
and compared these Lagrangians with those derived in \cite{IZ1,IZ3} within
the non-supersymmetric setting.

We gave a brief account of the  formalism of auxiliary superfields
for the ${\cal N}=1$ supersymmetric models with the $U(N)$ duality,
generalizing the similar formulation of the bosonic case \cite{IZ2}.
A few examples of the $U(N)$ self-dual models were
presented.

As for further perspectives, it seems important to extend the formulation with
auxiliary superfields  to the more general case with the
$Sp(2N,R)$ duality symmetry supported by the additional scalar chiral
superfields living in the coset $Sp(2N,R)/U(N)\,$. Also it would be
interesting to elaborate on the ${\cal N}=1$ version
of the proposal of Ref. \cite{IZ3}
about the possibility to deal, at all steps including quantization, with the
off-shell auxiliary (super)field representation
of self-dual (super)electrodynamics without  explicitly eliminating these
auxiliary objects.

\section*{Acknowledgements}
We thank Sergei Kuzenko for valuable correspondence.
We acknowledge a partial support from the RFBR grants Nr.12-02-00517
and Nr.13-02-91330, the grant DFG LE 838/12-1 and a grant of the
Heisenberg-Landau program.
E.I. \& B.Z. express their gratitude to the Institute of Theoretical Physics
of the Leibniz Universit\"at Hannover
for the kind hospitality in the course of this work.

\setcounter{equation}0

\end{document}